\documentclass{emulateapj}

\newcommand{\msun}{M$_{\sun}$\,}

\newcommand{\lsun}{L$_{\sun}$\,}

\newcommand{\mbylk}{${M/L_{K}}$}

\shorttitle{Stellar disk of M\,82}
\shortauthors{Mayya et al.}

\begin{document}

\title
{The Star Formation History of the Disk of the Starburst galaxy M\,82}

\author{Y. D. Mayya\altaffilmark{1}, A. Bressan\altaffilmark{1,2,3},
L. Carrasco\altaffilmark{1},  and L. Hernandez-Martinez\altaffilmark{4} }
\altaffiltext{1}
{Instituto Nacional de Astrofisica, Optica y Electronica, Luis Enrique Erro 1,
Tonantzintla, C.P. 72840, Puebla, Mexico}
\altaffiltext{2}
{Osservatorio Astronomico di Padova, Vicolo dell'Osservatorio 535122, Padova, 
Italy}
\altaffiltext{3}{SISSA, via Beirut 4, 34014, Trieste, Italy}
\altaffiltext{4}{Insituto de Astron\'omia, Universidad Nacional Autonoma de 
M\'exico, Apdo. Postal 70-264, Cd. Unversitaria 04510 México DF, Mexico}
\email{ydm@inaoep.mx}

\begin{abstract}
Spectroscopic, photometric and dynamical data of the inner 3~kpc part of the 
starburst galaxy M\,82 are analyzed in order to investigate the star 
formation history of its disk. The long-slit spectra along the major 
axis are dominated by Balmer absorption lines in the region outside the 
nuclear starburst all the way up to $\approx3.5$ scalelengths 
($\mu_B=22$~mag\,arcsec$^{-2}$). Single Stellar Population (SSP) spectra of 
age 0.4--1.0~Gyr match well the observed spectra in the 1--3~kpc zone, 
with a mean age of the stellar population marginally higher in the outer 
parts. The mass in these populations, along with that in the gas component, 
make up for the inferred dynamical mass in the same annular zone for a Kroupa 
initial mass function, with a low mass cut-off $m_l=0.4$~\msun.
The observed ratio of the abundances of $\alpha$ elements with 
respect to Fe, is also consistent with the idea that almost all the stars 
in M\,82 disk formed in a burst of short duration (0.3~Gyr) around 0.8~Gyr ago.
We find that the optical/near infrared colors and their gradients in the disk 
are determined by the reddening with visual extinction exceeding 1~mag even 
in the outer parts of the disk, where there is apparently no current star 
formation.
The disk-wide starburst activity was most likely triggered by the interaction 
of M\,82 with its massive neighbor M\,81 around 1~Gyr ago.
The properties of the disk of M\,82 very much resemble the properties of the 
disks of luminous compact blue galaxies seen at 0.2--1.0 redshift.

\end{abstract}


\section{Introduction}

M\,82 is a nearby edge-on galaxy (D = 3.63 Mpc, image scale
17.6~pc\,arcsec$^{-1}$; Freedman et al. 1994), which harbors a nuclear 
starburst, the prototype of the starburst phenomenon \citep{Riek80,Riek93}.
The optical appearance of M\,82 is dominated by bright star-forming knots
interspersed by dusty filaments \citep{OCon78}. This appearance
has led to an Irr II morphological classification of M\,82 \citep{Holm50}.
A near-infrared (NIR) bar of $\sim$1~kpc length was discovered by 
\citet{Tele91}. 
Recently, \citet{Mayy05} have discovered two symmetric spiral arms,
suggesting a morphological classification SBc.
Embedded inside the 500~pc starburst region is a small bulge, with an upper
mass limit of $3\times10^7$\msun \citep{Gaff93}. The $K$-band scalelength 
of the disk is as small as 860~pc \citep{Ichi95}.

M\,82 is a low mass galaxy ($\sim10^{10}$\,\msun), with a strong central 
concentration of mass and hardly any evidence for the dark matter halo
\citep{Sofu98}. This mass is in the lower mass range of late-type galaxies. 
However, the observed luminosity, mean surface brightness and metallicity are 
higher than expected for a late-type galaxy.
In addition, the percentage of mass in the gaseous component is as much as
$\sim50$\%, which is abnormally high for a normal galaxy \citep{Youn84}.
Several authors have attributed these abnormalities to some phenomenon 
associated with its interaction with other members of the M\,81 group. 
\citet{Sofu98} explained the small size of the disk and the falling rotation
curve with a model wherein M\,82 lost its outer disk and 
halo during a closeby interaction with M\,81 and NGC\,3077, which started 
around 1~Gyr ago.
\citet{Yun94} reproduced the observed structure of the H\,I streamers using
a numerical simulation, where M\,82 is interacting with M\,81 and NGC\,3077.
The closest approach occurred around 0.3~Gyr ago.
\citet{Elvi72} presented a picture wherein M\,82 captured a gas cloud
during its passage in the M\,81 group, in order to explain the high
gas fraction. Are the observed properties in the disk of M\,82, especially,
its high surface brightness, high metallicity, and low mass-to-light 
ratios, consistent with these scenarios? In this paper, we investigate 
the star formation history (SFH) that is consistent with the observed 
properties of the stellar disk, and compare the
resultant SFH with phenomena that may be related to the interaction.

In Section 2, we describe the spectroscopic and imaging observations used
in this work. Section~3 deals with a discussion of the quantities that 
can be used as constraints to the age of the stellar populations in the
disk of M\,82. In Section 4, we discuss the star formation and
chemical evolutionary histories that are consistent with our data.
The evolutionary status of M\,82 is discussed in Section 5. 
A summary of the main results of this work is given in Section~6. 

\section{Observations and Analysis} 

\subsection{Spectroscopic Observations}  

Spectroscopic observations of M\,82 were carried out with the purpose of
registering the absorption-line dominated stellar spectra at a variety of
radial distances from the center. Long-slit (3$^{\prime}$ in length) spectra 
were taken at three positions along the major axis of the galaxy 
(P.A. of the slit = $62^\circ$) using the Boller \& Chivens spectrograph
at the Guillermo Haro Astrophysical Observatory, Cananea, Mexico.
The observations were carried out on three nights, one in 1999 February and the
other two in 2003 November. The starburst nucleus was placed at the center,
extreme southwest and extreme northeast of the slit, respectively on the three
nights. With the three slit positions, spectra up to a major axis distance
of 3$^{\prime}$ were registered on either side of the nucleus. At this radial
distance, $B$-band surface brightness levels correspond to 
22~mag\,arcsec$^{-2}$. Two spectra, each of 30 minute duration, were taken for 
the bright central part, whereas three 30-minute spectra, were taken for the
other two slit positions. 
A slit width of $1.6^{\prime\prime}$ was employed.
A grating of 150~lines/mm was used which covers a wavelength range
of $\approx$3500--6800\,\AA\ at a spectral resolution of around 10\,\AA.
The spectral and spatial samplings were 3.2\,\AA/pix and 0.46\arcsec/pix
respectively.
Object was observed as close to the meridian as possible, with maximum
airmass of $\approx1.30$.
The instrumental response was calibrated by the observation of standard
stars HR\,1544, HR\,5501, BD+40 4032, Feige 15 and Feige 34.

Each frame was bias-corrected and divided by a normalized flat-field
using the tasks in the IRAF package. Wavelength-calibrated frames of the
same slit position were averaged, in the process removing cosmic ray events.
The disk of M\,82 occupies the entire slit-length and hence sky spectrum
could not be extracted from the same spectrum as the object.
Instead, sky spectra were extracted from the spectra of galaxies with
a compact nucleus observed immediately before and after the M\,82 
observations. 

\subsection{NIR and Optical Images} 

NIR images in the $J,H$ and $K$ bands used in the present work were taken
with the CAnanea Near Infrared CAmera (CANICA) 
\footnote{see http://www.inaoep.mx/\~{ }astrofi/cananea/canica/}
available at the 2.1-m telescope of the Observatorio
Astrof\'{\i}sico Guillermo Haro in Cananea, Sonora, Mexico. 
Details of these images
and the analysis procedure adopted were described in \citet{Mayy05}.
Optical images in the $U, B,V$ and $R$ bands are based on the data archived 
at the NASA Extragalactic Database.
These images were taken at the 40-inch telescope of the Mt. 
Laguna Observatory and were described by \citet{Marc01}.

\section{Constraints on the disk parameters from Observational data}

In this section, we combine our spectroscopic and photometric data with
published dynamical data to put strong constraints on critical parameters
of the stellar disk of M\,82.

\subsection{Spectroscopic ages of the stellar populations}

\begin{figure}[b]
\epsscale{1.3}
\plotone{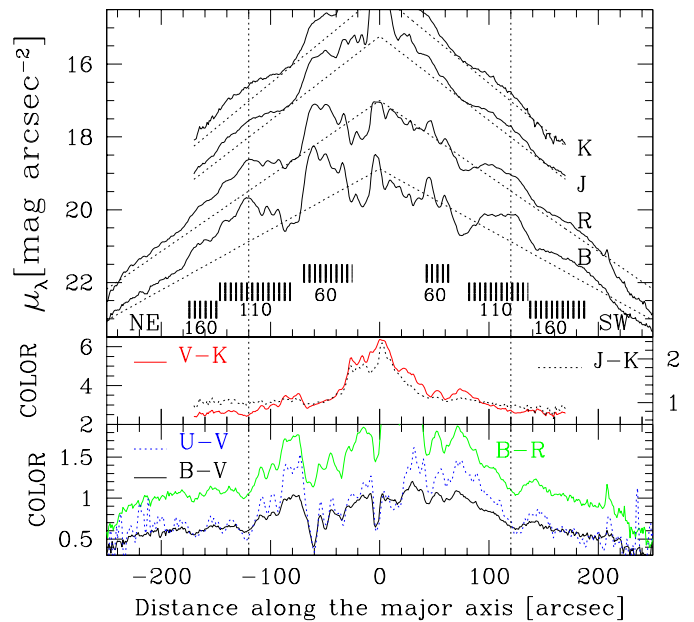}
\caption[]{One-dimensional intensity and color profiles of M\,82.
In the top panel, we show the intensity cuts along the major axis 
in $K, J, R$, and $B$ bands (solid lines), along with the exponential function
that fits best the azimuthally averaged intensity profiles in the
corresponding bands. The positions of the extracted spectra are shown
by vertical grids, with the associated numbers denoting the mean distance
in arcsecs of extracted regions from the nucleus. The NE and SW directions 
are also shown. The bottom two panels depict the major axis color profiles. 
The vertical 
dotted lines denote the intersection of the spiral arm with the major axis.
}
\end{figure}
Two-dimensional spectra of M\,82 were closely examined to isolate all regions
with noticeably distinct surface brightness values. On either side of the 
starburst nucleus, there are bright continuum emitting knots up to around 
60\arcsec\ (1~kpc) distance (e.g. region B on the northeast and region G
on the southwest as defined by O'Connell \&  Mangano 1978).
Spectra representative of these continuum knots were extracted on the 
northeast and 
southwest directions separately. Beyond a radius of 1~kpc on either side of the 
nucleus, bright knots are absent, and the spectra represent that of the 
underlying smooth disk. We defined two spectra on either side of the nucleus 
corresponding to radial zones between 1--2~kpc, and 2--2.7~kpc. 
The positions along the major axis of the extracted spectra are identified in 
Figure~1 by vertical hatches, which are labeled by their mean distances 
in arcsecs from the starburst nucleus. In this figure, we also show the major 
axis intensity cuts in the optical and NIR bands (solid lines). 
The exponential disk profiles (obtained by fitting the azimuthally averaged 
intensity profiles) are also plotted in this figure (dashed lines).
The bottom two panels show the major axis color profiles. The dashed vertical
lines mark the position of brightest region of the recently-discovered 
spiral arms, where the colors are marginally bluer. It is easily noticeable
that colors become systematically redder towards the center of the galaxy. 
\begin{figure}
\epsscale{1.30}
\plotone{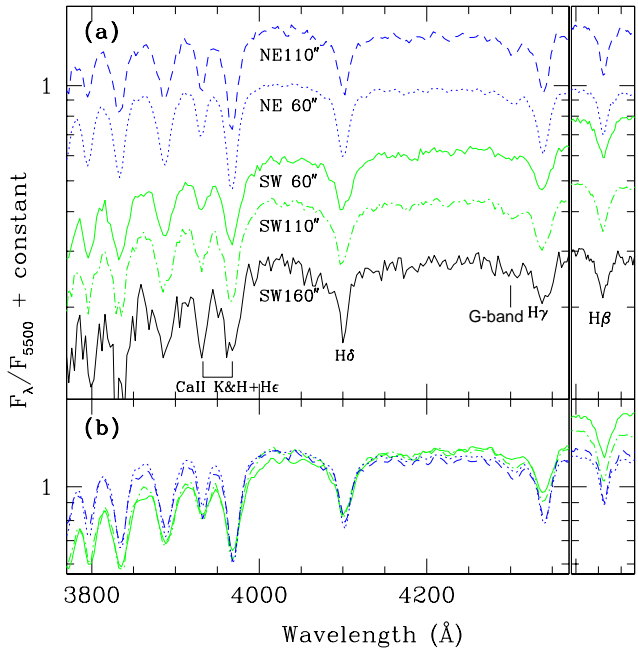}
\caption[]{Major-axis spectra of the stellar disk of M\,82 at five selected
zones. In (a), the top two spectra correspond to the northeast (NE), and the
bottom three correspond to the southwest (SW) part of the disk. The numbers
appearing after the NE or SW correspond to the radial distance from the 
nucleus in arcsecs of the extracted spectra.
Prominent age-sensitive absorption lines are identified. Note that the almost
feature-less part of the spectra between H$\gamma$ and H$\beta$ is not shown.
(b) The 60\arcsec\ and 110\arcsec\ spectra from either side are superposed
(line types and colors the same as in panel a). 
Note that the strengths of all the absorption features in the four spectra
are nearly identical, despite the NE spectra being systematically bluer 
than the SW spectra.
}
\end{figure}

The most critical part of the spectrum for population synthesis analysis 
is the blue part of the spectrum, because of the presence of a variety of
spectral features sensitive to young, intermediate-age and old populations.
This part of the extracted spectra are plotted in Figure~2. 
The signal-to-noise ratio of these spectra are good enough to 
measure spectral indices sensitive to age and metallicity.
Some of the important absorption lines are marked in the figure. In general, 
all spectra are dominated by Balmer absorption lines, which is consistent 
with the ``A--F'' spectral type inferred by \citet{OCon78}. 
Relative depths of the spectral lines in the four spectra corresponding to 
the regions in the inner disk are identical.  
This is illustrated in panel (b) of the figure where these four
spectra are superposed. The only difference is in the continuum shape, with
the NE spectra systematically bluer than the SW spectra. 
In comparison, the spectrum of the outer disk (SW110\arcsec) departs from the
inner disk spectra in having deeper Ca\,II K line as compared to the
Ca\,II H$+$H$\epsilon$ and other Balmer lines.

\begin{figure}
\epsscale{1.30}
\plotone{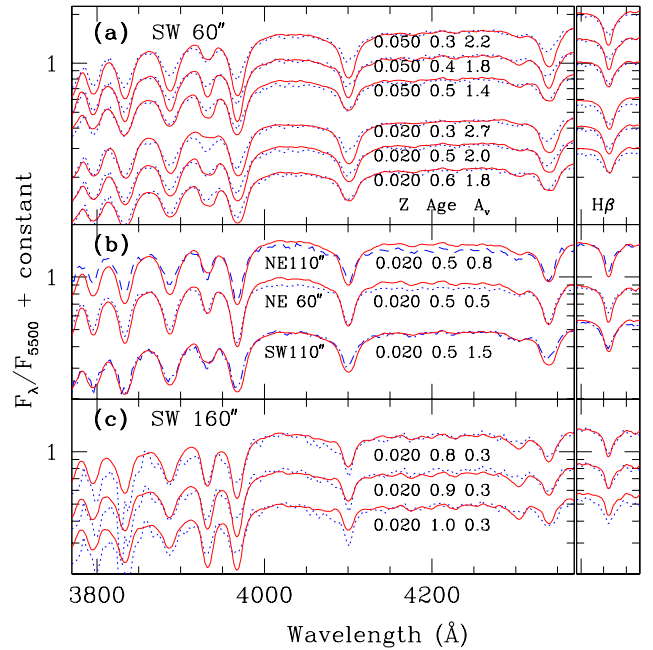}
\caption[]{
Observed spectra (dotted blue line) compared with model SSP spectra
(solid red line).
(a) Six models are superposed on the observed SW60\arcsec\ spectrum. 
The models correspond to two metallicities
(Z=0.050, 0.020), and three ages (0.3--0.6~Gyr). 
The best-match is obtained for 0.5~Gyr at Z=0.020.
(b) The best-match model SSP superposed on the rest of the inner disk spectra.
All these three spectra are consistent with an SSP of 0.5~Gyr at Z=0.020,
with the only difference between the spectra being the amount of reddening.
(c) Three models are superposed on the SW 160\arcsec\ spectrum.
This spectrum differs from the inner disk spectra in having the age-sensitive
Ca II\,K profile almost as deep as the Ca II\,H+H$\epsilon$
profile, an indication of relatively older population. The model that best 
matches the important observed features is at Z=0.020 for 0.9~Gyr. 
}
\end{figure}
In the remaining part of this section, we use the observed spectra to infer 
the age of the stellar populations that give rise to these features, using 
model spectra for Single Stellar Populations (SSPs), properly smoothed to 
the observed resolution. The model SSPs have a Kroupa like initial mass 
function (IMF) --- with masses between 0.15 and 120\,\msun\ 
\citep{1993MNRAS.262..545K}.  They are based on the Padova 
\citep{1993A&AS..100..647B, 1994A&AS..105...29F, 1994A&AS..104..365F}
evolutionary tracks and on the \citet{1984ApJS...56..257J}
library of stellar spectra. More details of the model can be found in 
\citet{1998A&A...332..135B}. It is important to stress here
that while we compare observed spectra with SSP models computed
with the usual Kroupa IMF, in Sec.~4 we will discuss also a top-heavy 
modification of the IMF. The use of such a top-heavy IMF does not change
our estimate of the age of the disk of M82.

Reddened model spectra are superposed on the observed spectrum for the
SW60\arcsec\ zone in Figure 3a, where we have made use of 
\citet{1989ApJ...345..245C} reddening curve.
The metallicity, age and the visual extinction of each model is indicated
below each displayed spectrum.
The intensity ratios of different features in the observed spectrum of the
inner disk can be reproduced only for ages between 0.3 and 0.6~Gyr, with the
exact age depending on the choice of metallicity. For ages
younger than 0.3~Gyr, the Ca II K line does not appear in the SSP, whereas 
for older ages it is much stronger than the observed value. 
The model that best matches the observed spectrum corresponds to an age of
0.5~Gyr and solar metallicity.
In Figure~3b, we illustrate that all the spectra in the inner disk 
be reproduced at solar metallicity for an age of 
0.5~Gyr, with extinction being different at different positions of the disk. 
Clearly, the southwestern side of
the disk is more reddened as compared to the northeastern side.
Derived ages would be younger (older) by around 0.1~Gyr if the metallicities 
are higher (lower) than solar values. More importantly, our spectra of inner 
disk regions are not consistent with mean ages greater than 0.7~Gyr.

The outer disk spectrum
suggests a population that is marginally more evolved as compared
to that of the inner disk. Nevertheless, the 4000~\AA\ break, which is a 
signature of populations a few gigayears old, is absent. We estimate an age 
of $0.9\pm0.1$~Gyr for the outer disk. In principle, acceptable fits can be
obtained by adding an old population of several gigayears, typical of galactic
disks, to a population of 0.5~Gyr old. However, single SSPs give a better 
match to the observed spectrum than those obtained by adding an underlying
old population.

As noted earlier, the NE60\arcsec\ spectrum corresponds to the the fossil
starburst region B.
The Hubble Space Telescope (HST) images of region B have revealed that this 
region consists of more than hundred super star clusters (SSCs).
\citet{deGr01} have estimated a mean age of these SSCs to be $0.65\pm0.25$~Gyr, 
with none of the clusters younger than 0.30~Gyr. Typical visual extinction
values were found to be less than 1~mag. 
These ages were estimated based on the optical-NIR (BVIJH) spectral energy 
distributions of individual clusters, keeping the reddening E(B-V) as a free 
parameter. Our spectroscopically estimated age of $0.5\pm0.1$~Gyr,
and $A_v=0.5$~mag are very much consistent with these photometrically estimated 
values. However, a re-analysis of the same HST dataset has yielded 
older ages $1.10\pm0.25$~Gyr \citep{deGr03}, but with mean $A_v<0.2$~mag.
It is well-known that both ageing and reddening have the same effect on the 
observable photometric quantities, and hence the age and extinction
cannot be derived completely independent of each other. On the other hand,
in the spectroscopic method that we have adopted, the relative strengths of 
absorption lines fix the age, whereas the slope of the continuum fixes the 
extinction, thus the age, and extinction were determined independent of each 
other. We re-iterate that the continuum of our spectra are consistent with 
both the solutions (low age with high extinction or high age with low 
extinction), but the absorption-line ratios clearly rule out ages exceeding 
0.7~Gyr for region B.

The similarity of the inner disk spectra, irrespective of whether they belong
to compact knots of region B or diffuse parts of the disk, carries important 
clues on the star formation scenario in the disk of M\,82.
In the next section, we use the mass-to-light ratio in the $K$-band to 
establish firm constraints on the mass of the underlying
old disk population, and also discuss a scenario of star formation in the
disk of M\,82.

\subsection{Absolute magnitude and colors}
\begin{figure}
\epsscale{1.3}
\plotone{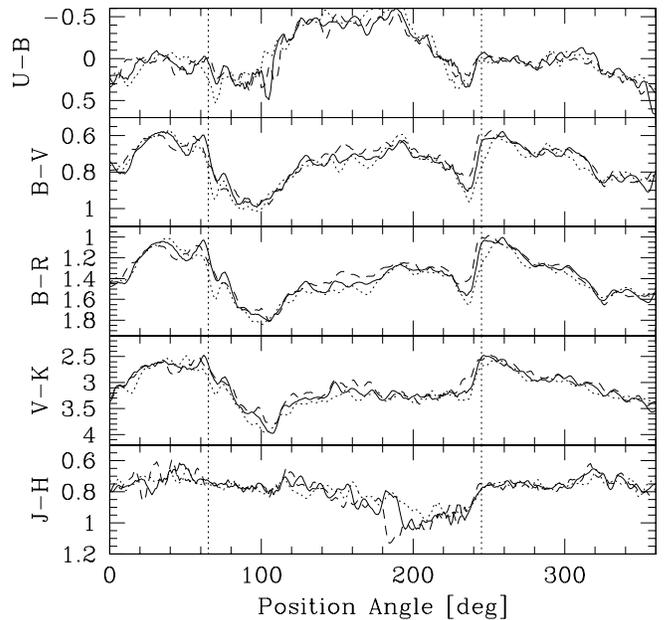}
\caption[]{
Azimuthal color profiles at radius=110, 120 and 130\arcsec\ from the center.
The position of the spiral arms is marked by the vertical lines. The arms,
in general, are bluer than the disk. The disk colors in the azimuthal
zone between 65--245$^\circ$ is affected by the foreground dust and the
superwind cone, and hence typical disk colors and their errors are derived 
averaging the colors between P.A. 320$^\circ$  and 360$^\circ$. 
}
\end{figure}

Azimuthally averaged intensity profiles of M\,82 in the NIR and the optical 
bands follow an exponential law expected for stellar disks of galaxies
(see Figure~1).
We generated the 2-dimensional exponential disk of M\,82 in the $K$-band
based on the observed scalelength of $47$\arcsec ($825$~pc).
We then extracted the disk magnitude in the annular zone between 70\arcsec\ 
and 170\arcsec, which was found to be 5.84~mag. At the distance of M\,82, 
this corresponds to an absolute magnitude of $-21.96$, without correction
for extinction.

Optical and NIR colors are useful in determining the evolutionary status of 
the stellar systems, if the reddening in various colors can be reliably 
estimated. The use of multiband images allows us to obtain profiles of 
azimuthal and radial distribution of colors. In Figure~4, we display the 
azimuthal distribution of colors in the radial zone between 90\arcsec\ and 
130\arcsec. In this radial zone, the arms are the strongest features, and 
hence these profiles can be used to investigate whether the results obtained 
from spiral-arm dominated major axis profile is applicable to the rest of 
the disk. In general, colors are the bluest along the major axis (dotted 
lines at P.A.s 65 and 245$^\circ$). However, the color difference between 
the disk and the spiral arms at a given radius is small compared
to the overall radial gradient in colors, displayed in Figure~1. Hence 
it seems that the results obtained from major axis spectra are representative 
of the entire disk. 

\subsection{The Mass and mass-to-light ratio}

The rotation curve of M\,82 has been derived in several studies using both
stellar and gaseous tracers up to a radius of 170\arcsec, with the 
intention of mapping the mass distribution in its disk. \citet{Maya60} 
and \citet{Gotz90} obtained a rotation curve using optical emission and
absorption lines, whereas \citet{Sofu98} used the CO and H\,I lines.
In all these studies, the rotation curve has been modeled to obtain the radial 
distribution of mass. \citet{Maya60} estimated a mass of $4\times10^9$\msun
in the 70--170\arcsec\ annulus, whereas \citet{Gotz90} masses are 50\% lower.
\citet{Sofu98} found that most of the mass is concentrated within the 
central 1~kpc radius, with the mass distribution outside this radius 
consistent with an exponential mass surface density profile. 
Significantly, there is no evidence for a dark matter halo, and hence
the entire mass outside the central bar can be associated with the stars
and gas in the disk.  We obtained the mass-to-light ratio
\footnote{Throughout this work, mass-to-light ratio is defined using the 
monochromatic $K$-band luminosity calibrated against the absolute magnitude 
of sun $M_{K\sun}=3.33$~mag. Instead, if we use the $K$-band 
luminosities integrated over a standard $K$-band response curve of band 
width 0.35$\mu$m, and express mass-to-light ratio in \msun/\lsun,
then the quoted values have to be multiplied by 50.}
of the disk outside a radius of 1~kpc by combining the disk mass 
and the $K$-band luminosity in the same annular zone.
The resulting value is $0.30\pm0.05\,{{M_{\sun}}/{L_{K\sun}}}$ if we 
use mass estimates of \citet{Maya60} and 0.15 for the mass estimates of 
\citet{Gotz90}.
At least 15\% of the disk mass in the annular zone outside 1~kpc is
in the form of atomic and molecular gas \citep{Youn84}.

\subsection{Metallicity of the nebular gas and stars}

Metallic abundances of gas and stars in the starburst region of M\,82
have been analyzed by \citet{Orig04}. The cold phase gas abundances were
derived using the optical and NIR nebular lines associated with the giant
HII regions, whereas the stellar abundances were obtained using the
NIR absorption lines that originate in the atmospheres of cool red supergiants.
Since these stars live only for a few tens of million years, both the stellar
and gaseous abundances are expected to represent the values present in the
disk before the onset of nuclear starburst.

In Table~1, we summarize all the observed quantities for the disk, which we 
will use as a constraint for a detailed modeling of
the star formation and chemical history of the disk of M\,82.

\section{Star formation and chemical history of the disk of M\,82} 

In order to reproduce the observed quantities (metallicity enhancement,
\mbylk, age and colors), we ran a number of chemical evolutionary models 
from the GRASIL WEB interface GALSYNTH 
\footnote{http://web.pd.astro.it/galsynth/index.php}.
In these models, gas phase metallicities of different elements are 
calculated over the Hubble time for a given SFH of the galaxy. At each epoch, 
the star formation rate (SFR) is calculated using a Schmidt law, with the 
exponent value of 1.0. The standard Kroupa IMF parameters are used
\citep{1993MNRAS.262..545K}. 
The current \mbylk\ value is calculated as the ratio of the mass locked up in 
stars (both living as well as their remnants) to the luminosity of all the 
living stars. It is important to note that in order to reproduce the observed 
\mbylk\  values, it is necessary to correct the model values for the mass in gas
($>15$\%), and the extinction in the $K$-band.

Among the observed quantities, the metallicity enhancement critically depends 
on the slope and the upper cut-off mass of the IMF and the burst duration, 
whereas the lower cut-off mass of the IMF and the age of the burst determine 
the \mbylk\  ratios and colors. The observed set of parameters could be produced 
only by those models where almost all the observed stellar mass of the disk 
was formed by a burst lasting for only a few hundreds of million years, the 
burst having completely stopped around 0.5~Gyr.
We ran a set of models for a wide range of input parameters. 
We discuss two of these models below.

The run of the SFR of the best-fit model as a function of the age of the 
living stars is shown in Figure~5a. 
In Figure~5b, we show the metallicity enhancements for various metals for 
the standard Kroupa IMF 
(open circles denoted model A), along with the observed 
enhancements in the cold gas and stars. The corresponding values for the
\mbylk\  and colors are shown in the bottom panel of the figure.
With the Kroupa IMF the model produces metallic enhancements which are 
systematically outside the observed values even after taking into account
the errors
in the observed metallicities. The model also produces \mbylk\  values
which are around 30\% higher than the highest values allowed by the 
observations. The model colors, after reddening them with the inferred
extinction values from the spectra, are consistent with the observations. 
\begin{center}
\begin{deluxetable}{llccccc}
\tablewidth{0pc}
\tablecaption{Observed and model properties of the disk of M\,82}
\tablehead{
\colhead{Quantity} & \colhead{units} & \colhead{Observed} 
                   & \colhead{Model A} & \colhead{Model B} \\
}
\startdata
$M_{\rm disk}$   & $10^9$\msun  
                         &  $<4.0$          & 4.0      & 4.0      \\
$M({\rm gas})\over{M({\rm disk})}$ 
                & \nodata&  $>0.15$         & 0.15     & 0.15  \\
$M_{\rm K}$(disk)& mag   &  $-21.96\pm0.05$ & $-21.42$ & $-21.95$ \\
$M/L_{\rm K}$    & \msun$/L_{\rm K_\sun}$  
                         &  $<0.30\pm0.05$  & 0.53     & 0.32  \\
$[O/Fe]$         &\nodata& $0.35\pm0.20$    & 0.11     & 0.21  \\
\hline
\enddata
\end{deluxetable}
\end{center}
The only way we could reproduce the observed high enhancements of $\alpha$ 
elements and a low \mbylk\  is by adopting a top-heavy IMF 
(model B). In this model, the slope of the Kroupa IMF for the massive stars 
is reduced from 1.7 to 1.4 (i.e. closer to the Salpeter slope), 
and the lower mass cut-off raised from 0.15\ to 0.4\,\msun. The resulting 
values are shown in Figure~5 by square symbols. This model successfully 
reproduces the observed trend in the metallic enhancements for all the 
elements except Carbon. The \mbylk\  values are also very well reproduced. In 
fact, the agreements between the model and
the observational values would be even better if we use a flatter IMF.
These trends for the disk IMF are very similar to that found for the 
nuclear IMF by \citet{Riek93}. 
The shallower IMF would imply a harder ultraviolet radiation field,
contrary to what is observed for a sample of 23 starburst nuclei
\citep{Rigby04}.
This apparent contradiction could be solved if the most massive stars spend
most of their lifetimes embedded in dense highly extinguished regions, as
suggested by \citet{Silva98}.
The model quantities are compared with the observed quantities in Table~1.

In summary, models with more or less constant SFR over the Hubble time cannot 
reproduce neither the observed low mean age of the disk of M\,82,
nor the observed chemical pattern and \mbylk\  values. On the other hand,
observations can be only reproduced if more than 90\% of the stellar mass 
inside the central 3~kpc is younger than a gigayear.
Hence, around 1~Gyr ago, the disk experienced an intense starburst.
Furthermore, we can reproduce both the low \mbylk\  value and the 
enhancement of the $\alpha$ elements only by a suitable modification
of the Kroupa IMF. To reproduce the observed \mbylk, half a gigayear after 
the burst has stopped, we are forced to truncate the lower end of the IMF at 
0.4 \msun. Moreover, in order to reproduce the observed enhancement with SNIa 
progenitors that originate from mass accretion onto white dwarfs in binary 
systems (as it is adopted in our chemical model) we need to decrease the 
slope of the IMF of the massive stars to the Salpeter value. 
Thus, by combining age, luminosity 
and dynamical mass of the stellar populations we conclude that the disk of 
M\,82 formed very recently and with an IMF peaked toward the massive stars.
The increase in the low-mass cutoff and decrease in the slope may be due to 
a turbulence-enhanced ISM, probably as a result of interaction or/and 
the starburst episode itself. 
\begin{figure}
\epsscale{1.3}
\plotone{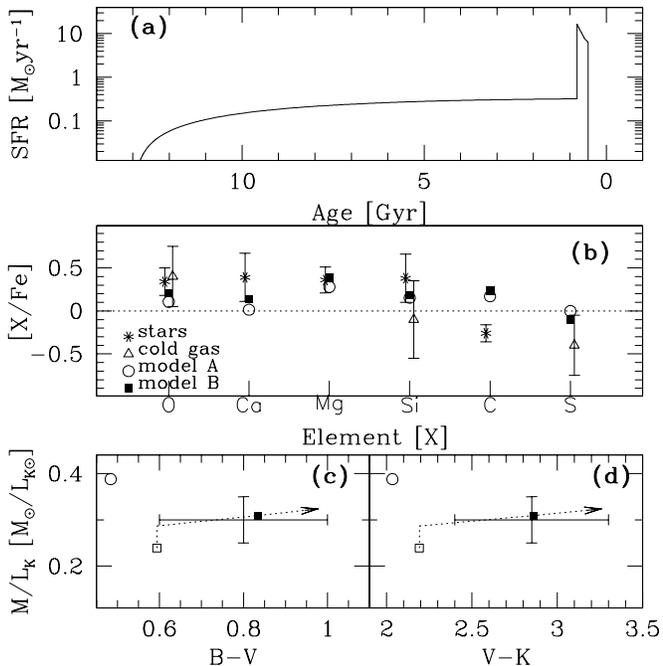}
\caption[]{(a) Star formation History of the disk of M\,82 that reproduces 
best the observed properties. 
(b) Observed stellar and nebular metallicities compared with the
gas metallicities of our best-fit model. Model A and B correspond to standard
and top-heavy Kroupa IMFs, respectively.
(c--d) $B-V$ and $V-K$ colors are
plotted against the K-band mass-to-light ratio of the disk.
The error bars on the colors are an indication of the observed color gradient
between 80--160\arcsec\ part of the disk. 
Open circles and squares correspond to Model A and B, 
respectively, without taking  into account the effect of the gas mass on the 
\mbylk\ ratio and the interstellar extinction on the colors. In both 
these models, the burst is of 0.3~Gyr duration, which started around
0.8~Gyr ago. Star formation activity in the disk has ceased since then.
The dotted paths indicate the locus of 
this point after corrections for gas content (fraction of gas mass is 15\%) 
and extinction (length of the arrow corresponds to $A_v=1.2$~mag) are applied. 
Model B successfully reproduces all the observed trends.
}
\end{figure}

\section{Evolutionary status of M\,82}

Disks of normal spiral galaxies are characterized by star formation 
that is continuing un-interrupted over the Hubble time 
\citep{1994ApJ...435...22K}.
M\,82 differs in two ways from these normal disks: (1) luminosity weighted
age of the disk is only 0.5~Gyr, which is an order of magnitude younger,
and (2) there is no present star formation in the disk. 
M\,82 also differs from the normal galaxies in the relations between
global parameters, it is too bright and also metal rich for its low mass
value \citep{2004AJ....127.2031K}. 
The origin of these peculiarities of M\,82 could be understood with our 
inferred SFH, in which almost the entire observed stellar mass was formed
in a short-duration burst, rather than in a continuous manner.
In our best-fit model, the disk-wide burst started around 0.8~Gyr 
ago, which is consistent with the interaction scenario proposed by 
\citet{Sofu98}. 
It is interesting to note that a number of interacting galaxies also
show disk-wide star formation at present: examples being NGC\,4038/9,
Arp\,299 and NGC\,4676 \citep{1998A&A...333L...1M, 1999AJ....118..162H,
2000ApJ...541..644X}. This kind of widespread star formation in galactic disks 
is probably induced by shocks generated in the interstellar medium following 
the interaction \citep{2004MNRAS.350..798B}.

\citet{Miho99} discussed pre-collisional galaxy properties that favors 
star formation in the disk, before the burst is turned on in the center.
He found that the pre-collisional galaxy should be a normal high-surface 
brightness galaxy with a big bulge or a gas-rich low surface brightness 
galaxy (LSB). 
If the pre-collisional M\,82 had a big bulge, it would have survived the
interaction \citep{Sofu98}. However, the observed bulge of M\,82 is small,
consequently, pre-collisional M\,82 should have been a gas-rich LSB galaxy. 
This is in agreement with our proposed SFH.
It could be that either M\,82 was a gas-rich LSB galaxy before the interaction, 
or that it acquired vast amounts of gas during the interaction. 
Below we discuss three scenarios that are consistent with the present
appearance of M\,82.

\subsection{Kinematically disturbed normal galaxy}

The most important distinguishing characteristic of M\,82 is its low value 
of M/L ratio as compared to that of normal galaxies. Is this because of 
an underestimation of its mass? The kinematics of the galaxy could have been
disturbed because of interaction resulting in possible underestimation of 
its mass. Blue-band spectra would look identical even if we add
upto a factor of 3--5 more mass in stars older than 5~Gyr as compared to 
the mass in stars 
10 times younger. However, the presence of such an amount of old stars
would have contributed to the enrichment of Fe, without the corresponding
enrichment of $\alpha$ elements, thus resulting in low values of [$\alpha$/Fe].
Thus the observed enhancement of $\alpha$ elements indicates that the galaxy
really lacks a significant population of stars older than 1~Gyr.

\subsection{Rejuvenated star formation in a stripped stellar disk}

In the scenario proposed by \citet{Sofu98}, much of the disk and halo of 
M\,82 were stripped off during its interaction with the members of the M\,81
group. However, none of the observational studies so far have detected the 
stripped dark matter halo, containing the old disk. It would be interesting 
to carry out a search for such a system in the M\,81 group.
Even in this scenario, the gas and stars in the central kiloparsec region
are not stripped, and hence the observed $\alpha$ enhancement of the nuclear 
gas requires vast amounts of star formation, and inflow of the 
metal-enriched material, after the original disk has been lost.
In other words, the SFH we have inferred is consistent
with this scenario, and hence 
M\,82 is in the process of re-building its stellar disk.

\subsection{A nearby galaxy in formation?}

Analysis of galaxies at redshifts up to 1 support a morphological evolution
over the past few gigayears. The catalyst that drives the morphological 
evolution is the major mergers between galaxies. During the merging process 
galaxies grow both in size and luminosity. Data suggest that at least 50\% 
of the nearby disk galaxies are formed through this scenario 
\citep{2005A&A...430..115H}.
The most important ensemble of galaxies that are thought to be caught 
in the formation process are the luminous blue compact galaxies (LBCGs)
seen between redshifts 0.2--1.0. The characteristic signature of these 
galaxies is the enhanced star formation in their nuclear regions. In a recent 
study, \citet{2006ApJ...640L.143N} have characterized the properties of the 
stellar disk surrounding the compact nucleus, and found that they are, in 
general, short in scalelength as compared to the scalelengths of local 
galaxies.

M\,82 resembles these LBCGs in having high luminosity, short scalelength, 
a nuclear starburst and low mass. Moreover, interaction is responsible
for all the observed characteristics in M\,82. Thus in M\,82, we may be
witnessing stars being formed for the first time in its disk, probably
in the debris left behind in the interaction. In other words, M\,82
could be a nearby galaxy in formation.

\section{Summary}

We used spectrophotometric, photometric and dynamical data of the 
inner $\approx$3\arcmin\ (3~kpc) radius of M\,82, to study the star formation 
history in the inner disk of M\,82.
The optical diameter (measured at $\mu_B=25$\,mag\,arcsec$^{-2}$) of M\,82 
is around 13$^\prime$, and hence our analysis pertains to the inner half
of the disk. We find that a large fraction of the 
inner disk is as young as 0.5~Gyr.
The models that best reproduce the above observations indicate
that the star formation in the disk happened in a burst of 0.3~Gyr 
duration, that ended around 0.5~Gyr ago. The short duration of the burst
is not enough to reproduce both the enhancement of the $\alpha$ elements and 
the low value of the \mbylk\ ratio, and a suitable top-heavy IMF must be invoked.
It is possible that the star formation started in the whole disk
about 0.8~Gyr ago, induced by shocks in the interstellar medium following the
interaction with members of M\,81 group, and now continues only in the 
central regions.

\acknowledgments

We are grateful to  L. Silva and G. L. Granato for providing us 
their chemical evolutionary model before its publication.
We thank Abelardo Mercado for carrying out some of the spectroscopic
observations used in this study. Suggestions by an anonymous referee have
greatly helped in improving the presentation of this manuscript.
This work was partly supported by the CONACyT (Mexico) projects 39714-F
and G28586E.
This research has made use of the NASA/IPAC Extragalactic Database, 
which is operated by the Jet Propulsion Laboratory, California
Institute of Technology, under contract with the National Aeronautics and Space
Administration.



\begin{thebibliography}{}

\bibitem[Barnes (2004)]{2004MNRAS.350..798B}
Barnes, J.~E.\ 2004, \mnras, 350, 798

\bibitem[Bressan et al.(1993)]{1993A&AS..100..647B} 
Bressan, A., Fagotto, F., Bertelli, G., \& Chiosi, C.\ 1993, \aaps, 100, 647

\bibitem[Bressan et al.(1998)]{1998A&A...332..135B} 
Bressan, A., Granato, G.~L., \& Silva, L.\ 1998, \aap, 332, 135

\bibitem[Cardelli et al.(1989)]{1989ApJ...345..245C} 
Cardelli, J.~A., Clayton, G.~C., \& Mathis, J.~S.\ 1989, \apj, 345, 245

\bibitem[Elvius (1972)]{Elvi72}
Elvius, A. 1972, \aa, 19, 193

\bibitem[Fagotto et al.(1994a)]{1994A&AS..105...29F} 
Fagotto, F., Bressan, A., Bertelli, G., \& Chiosi, C.\ 1994a, \aaps, 105, 29

\bibitem[Fagotto et al.(1994b)]{1994A&AS..104..365F} 
Fagotto, F., Bressan, A., Bertelli, G., \& Chiosi, C.\ 1994b, \aaps, 104, 365

\bibitem[Freedman et al. (1994)]{Free94}
Freedman, W. L. et al. 1994, \apj, 427, 628

\bibitem[Gaffney, Lester \& Telesco (1993)]{Gaff93}
Gaffney, N. I, Lester, D. F,  \& Telesco, C. M. 1993, \apjl, 407, 57

\bibitem[Goetz et al.(1990)]{Gotz90} 
Goetz, M., McKeith, C.D., Downes, D., \& Greve, A. 1990, \aap, 240, 52

\bibitem[de Grijs, O'Connell \& Gallagher(2001)]{deGr01}
de Grijs, R., O'Connell, R. W., \& Gallagher, J. S. 2001, \aj, 121, 768

\bibitem[de Grijs, Bastian \& Lamers(2003)]{deGr03}
de Grijs, R., Bastian, N., \& Lamers, H. J. G. L. M. 2003, \mnras, 340, 197

\bibitem[Hammer et al.(2005)]{2005A&A...430..115H} 
Hammer, F., Flores, H., Elbaz, D., Zheng, X.~Z., Liang, Y.~C., 
\& Cesarsky, C.\ 2005, \aap, 430, 115

\bibitem[Hibbard \& Yun(1999)]{1999AJ....118..162H}
Hibbard, J.~E., \& Yun, M.~S.\ 1999, \aj, 118, 162

\bibitem[Holmberg (1950)]{Holm50}
Holmberg, E. 1950, Lund Medd. Astron. Obs. Ser. II, 128, 1

\bibitem[Ichikawa et al. (1995)]{Ichi95}
Ichikawa, T., Yanagisawa, K., Itoh, N., Tarusawa, K., van Driel, W., 
\& Ueno, M.\ 1995, \aj, 109, 2038 

\bibitem[Jacoby et al.(1984)]{1984ApJS...56..257J}
Jacoby, G.~H., Hunter, D.~A., \& Christian, C.~A.\ 1984, \apjs, 56, 257

\bibitem[Karachentsev et al.(2004)]{2004AJ....127.2031K}
Karachentsev, I.~D., Karachentseva, V.~E., Huchtmeier, W.~K., \&
Makarov, D.~I.\ 2004, \aj, 127, 2031

\bibitem[Kennicutt et al.(1994)]{1994ApJ...435...22K} 
Kennicutt, R.~C., Tamblyn, P., \& Congdon, C.~E.\ 1994, \apj, 435, 22

\bibitem[Kroupa et al.(1993)]{1993MNRAS.262..545K} 
Kroupa, P., Tout, C.~A., \& Gilmore, G.\ 1993, \mnras, 262, 545 

\bibitem[Marcum et al.(2001)]{Marc01}
Marcum, P.~M., et al.\ 2001, \apjs, 132, 129 

\bibitem[Mayall (1960)]{Maya60}
Mayall, N. U. 1960, Ann. d'Ap., 23, 344

\bibitem[Mayya, Carrasco \& Luna (2005)]{Mayy05}
Mayya, Y.~D., Carrasco, L., \& Luna, A.\ 2005, \apjl, 628, L33 

\bibitem[Mihos(1999)]{Miho99}
Mihos, J.~C.\ 1999, IAU Symp.~186: Galaxy Interactions at Low and High 
Redshift, 186, 205

\bibitem[Mirabel et al.(1998)]{1998A&A...333L...1M}
Mirabel, I.~F., et al.\ 1998, \aap, 333, L1

\bibitem[Noeske et al.(2006)]{2006ApJ...640L.143N} 
Noeske, K.~G., Koo, D.~C., Phillips, A.~C., Willmer, C.~N.~A., 
Melbourne, J., Gil de Paz, A., \& Papaderos, P.\ 2006, \apjl, 640, L143

\bibitem[O'Connell \& Mangano (1978)]{OCon78}
O'Connell, R. W., \& Mangano, J. J. 1978, \apj, 221, 62

\bibitem[Origlia et al.(2004)]{Orig04}
Origlia, L., Ranalli, P., Comastri, A., \& Maiolino, R. 2004, \apj, 606, 862

\bibitem[Rieke et al. (1980)]{Riek80}
Rieke, G. H., Lebofsky, M. J., Thompson, R. I., Low, F. J., \& Tokunaga, A. T.
1980, \apj, 238, 24

\bibitem[Rieke et al.(1993)]{Riek93} 
Rieke, G.~H., Loken, K., Rieke, M.~J., \& Tamblyn, P.\ 1993, \apj, 412, 99

\bibitem[Rigby \& Rieke (2004)]{Rigby04}
Rigby, J.~R. \& Rieke, G.~H. 2004, \apj, 606, 237 

\bibitem[Silva et al. (1998)]{Silva98}
Silva, L., Granato, G. L., Bressan, A., \& Danese, L. 1998, ApJ, 509, 103

\bibitem[Sofue (1998)]{Sofu98}
Sofue, Y. 1998, \pasj, 50, 227

\bibitem[Telesco et al. (1991)]{Tele91}
Telesco, C. M., Joy, M., Dietz, K., Decher, R., \& Campins, H. 1991,
\apj, 369, 135

\bibitem[Young \& Scoville (1984)]{Youn84}
Young, J.S., \& Scoville, N. Z. 1984, \apj, 287, 153

\bibitem[Yun, Ho, \& Lo (1994)]{Yun94}
Yun, Min S., Ho, Paul T. P., \& Lo, K. Y. 1994, Nature, 372. 530

\bibitem[Xu et al.(2000)]{2000ApJ...541..644X}
Xu, C., Gao, Y., Mazzarella,
J., Lu, N., Sulentic, J.~W., \& Domingue, D.~L.\ 2000, \apj, 541, 644

\end{thebibliography}
\end{document}